\documentclass[letterpaper]{article} 
\usepackage{algorithm}
\usepackage{algorithmic}
\usepackage{xcolor}

\usepackage{booktabs,cleveref}
\usepackage{tabularx,array}
\usepackage{makecell}
\usepackage{ragged2e}
\newcolumntype{P}[1]{>{\RaggedRight\arraybackslash}p{#1}}

\usepackage{aaai25}  
\usepackage{times}  
\usepackage{helvet}  
\usepackage{courier}  
\usepackage[hyphens]{url}  
\usepackage{graphicx} 
\usepackage{natbib}  
\usepackage{caption} 
\usepackage{algorithm}
\usepackage{algorithmic}
\nocopyright
\usepackage{newfloat}
\usepackage{listings}
\DeclareCaptionStyle{ruled}{labelfont=normalfont,labelsep=colon,strut=off} 
\floatstyle{ruled}
\newfloat{listing}{tb}{lst}{}
\floatname{listing}{Listing}
\title{Context-Aware Detection and Victim-Centered Response Generation for Online Harassment in Private Messaging}
\author {
    Pinxian Lu\textsuperscript{\rm 1},
    Nimra Ishfaq\textsuperscript{\rm 2},
    Emma Win\textsuperscript{\rm 3},
    Morgan Rose\textsuperscript{\rm 3},
    Sierra R Strickland\textsuperscript{\rm 3},
    Candice L Biernesser\textsuperscript{\rm 3},
    Jamie Zelazny\textsuperscript{\rm 3},
    Munmun De Choudhury\textsuperscript{\rm 1},
}
\affiliations {
    \textsuperscript{\rm 1}Georgia Institute of Technology\\
    \textsuperscript{\rm 2}The University of Texas at Austin\\
    \textsuperscript{\rm 3}University of Pittsburgh\\
    plu74@gatech.edu, nimraishfaq@utexas.edu, wine@upmc.edu, rosem5@upmc.edu, stricklandsr3@upmc.edu, lubbertcl@upmc.edu, jmz22@pitt.edu, munmun.choudhury@cc.gatech.edu,
}

\usepackage{bibentry}

\begin{document}

\maketitle

\begin{abstract}
Online harassment is a widespread social and public health concern, yet most computational approaches for detecting and addressing harassment focus on publicly visible social media content rather than private messaging environments. Private conversations present unique challenges because harmful interactions often unfold through context-dependent, multi-turn exchanges, while victims may lack timely support during moments of harassment. In this study, we investigate how large language models (LLMs) can support both the detection of and response to online harassment in private messaging. Using a dataset of 80,053 Instagram direct messages donated by 26 adolescents aged 12-18, including youth with suicide risk factors, we first construct a human-labeled dataset of online harassment in private conversations and develop a context-aware cascading LLM classification pipeline. The proposed pipeline outperforms baseline toxicity classifiers trained primarily on public social media data. We then develop a victim-centered response framework that produces context-sensitive and psychologically-grounded AI-generated responses to online harassment messages. Human evaluators perceived the AI-generated responses as significantly more helpful than the original participant responses (95\% CI: 0.767--0.815, $p < .001$), particularly in terms of emotional support and de-escalation. Our findings highlight the potential of context-aware and victim-centered AI systems to provide just-in-time support during harassment in private messaging environments.


\noindent \textit{Content Warning:} This paper includes mentions, descriptions, and examples of online harassment.
\end{abstract}

\section{Introduction}

While social media enables new forms of entertainment, communication, and social connection, it also creates conditions that can intensify harmful interpersonal behavior online. Platform features and affordances such as anonymity, asynchronicity, and the absence of social cues can reduce accountability and increase aggression, consistent with the online disinhibition effect, where individuals behave differently online than they do in person \cite{suler2004online, nesi2018transformation}. As a result, online harassment has emerged as a widespread public health and social concern 
\cite{vogels2021state}. 
Adolescents are particularly vulnerable: in 2025, 32.7\% of American adolescents reported experiencing cyberbullying in the previous 30 days 
\cite{patchin_2022}. Moreover, online harassment is associated with elevated risks of adverse mental health outcomes, including suicidal ideation, especially among adolescents \cite{nesi2021social}. These concerns may be particularly acute among youth already experiencing suicide risk factors, who may encounter more severe forms of online harm and may be especially vulnerable to the psychological consequences of cybervictimization. Understanding how to identify and support such youth during moments of online harassment is thus an important and understudied challenge.


To combat online harassment and other harmful content, substantial research has focused on developing automated detection systems and interventions. Researchers have used machine learning approaches to identify and label online harassment content \cite{bretschneider2014detecting,yin2009detection}, while interventions such as reconsideration prompts and counter-speech have also been proposed and evaluated \cite{katsaros2022reconsidering, hangartner2021empathy}. However, most of this work has centered on publicly visible social media content, despite evidence that online harassment frequently occurs in private messaging spaces and may involve more severe forms of abuse \cite{vogels_2021_1}. Private conversations often contain richer interpersonal context, escalating exchanges, and sensitive disclosures that are not observable in public-facing platforms, making them especially important settings for understanding and addressing online harassment. Moreover, unlike public posts, private messages often unfold through sustained, relational, and context-dependent exchanges, where the meaning and severity of harassment may only become visible across multiple conversational turns.

Yet, datasets concerning online harassment in private messaging remain largely unavailable. Existing research efforts predominantly rely on using publicly accessible social media posts  for harassment detection \cite{golbeck2017large,davidson2017automated}. 
Models trained primarily on isolated public posts may fail to capture the interpersonal dynamics, escalation patterns, and contextual nuances that characterize harassment in private messaging environments. 
Moreover, because identifying harassment in conversational settings requires sustained exposure to sensitive and harmful interactions across longer contexts, scalable methods that reduce annotator burden are important \cite{umarova2024xenophobia}.

In addition, existing interventions against online harassment remain limited, particularly in private messaging contexts. Social media platforms predominantly encourage users to document, report, block, or ignore abusive interactions \cite{help_center_2017, how_to_handle_x_abuse, facebook_help_center}. However, these approaches are often reactive rather than supportive: they place the burden of action on victims after harm has already occurred, provide limited transparency into moderation outcomes, and may leave users with little sense of agency or emotional support \cite{vilk_lo_2023}. At the same time, many interventions focus primarily on modifying perpetrator behavior through content moderation, reconsideration prompts, or counterspeech \cite{katsaros2022reconsidering, hangartner2021empathy, bar2024generative}. While valuable, these approaches have shown mixed effectiveness and may not adequately support individuals experiencing harassment in the moment it occurs. This need for timely and supportive intervention may be especially important for adolescents with suicide risk factors, for whom experiences of online harassment may compound existing emotional distress and vulnerability 
\cite{quintana2022beyond}. Together, these gaps point to the need for clinically-grounded and theoretically-situated interventions that directly support victims during and immediately after online harassment.

Recent advances in large language models (LLMs) create new opportunities to address these limitations in online harassment detection and intervention. Unlike traditional moderation systems that primarily focus on flagging harmful content or penalizing perpetrators, LLMs can interpret conversational context, generate contextually appropriate language, and provide interactive forms of support in real time. Prior work has shown the promise of generated text in domains such as healthcare, peer support, and education, where LLMs have been used to enhance communication, empathy, and decision support \cite{zaretsky2024generative,lee2024large,wang2024tutor}. These capabilities are especially relevant in private messaging environments, where harassment often unfolds through nuanced, multi-turn exchanges and where victims may need immediate, context-sensitive support rather than delayed moderation outcomes. At the same time, LLMs may also help address challenges in large-scale harassment detection by leveraging conversational context without requiring extensive manually labeled datasets. Despite this potential, the use of LLMs to detect and respond to online harassment in private messaging contexts remains largely unexplored. Accordingly, our broader goal is to investigate how LLMs can support scalable, context-aware, and victim-centered approaches for detecting and responding to online harassment in private conversations.


Specifically, we address the following research questions: 

\begin{itemize}
    \item \textbf{RQ1:} \textit{How can we effectively identify online harassment in private messaging on a large scale?}
    \item \textbf{RQ2: }\textit{How can we utilize LLMs to help address online harassment in private messaging?}
\end{itemize}

To address these RQs, we analyzed a dataset of 80,053 Instagram direct messages donated by 26 adolescents aged 12-18, including youth with suicide risk factors. We first constructed a human-labeled dataset of online harassment in private conversations and developed an LLM-based classification pipeline that leverages conversational context to identify online harassment at scale. We evaluated this pipeline against two baseline classifiers and found that it achieved stronger performance in detecting online harassment in private messaging contexts. We then developed an LLM-based response generation pipeline to produce context-sensitive responses to online harassment messages. Human evaluators compared these AI-generated responses against the adolescents’ original responses and perceived the AI-generated responses as significantly more helpful overall, particularly in supporting de-escalation and emotional coping.

Together, our findings demonstrate the potential of LLMs to support scalable, context-aware, and victim-centered approaches for detecting and responding to online harassment in private messaging environments. Broadly, this work points toward the design of just-in-time AI-mediated interventions that can support vulnerable youth during moments of online harm.

\section{Related Work}

\subsection{Detection of Online Harassment}

Nearly a decade of prior research on online harassment detection has developed a range of computational approaches, including rules-based systems, traditional machine learning, and pattern-based methods \cite{mahbub2021detection, yin2009detection, bretschneider2014detecting}. Subsequent work has explored unsupervised learning, ensemble methods, and deep learning models for detecting harmful or abusive content across social media platforms \cite{di2016unsupervised, azeez2023classification, akhter2022abusive, anand2019classification, kumbale2023bree}. More recent work has also begun to examine harmful interactions in Instagram conversations using multimodal methods \cite{ali2023getting,alsoubai2025timeliness}.

However, most harassment detection work remains oriented toward public or semi-public social media content, where individual posts or comments are often treated as the primary unit of analysis. Private messaging creates a different detection problem: harmful meaning may depend on prior conversational turns, relationship dynamics, sarcasm, escalation, or repeated exchanges that are not visible in isolated messages. This distinction is especially important because private conversations are substantially less accessible to researchers and platforms, limiting opportunities to study online harassment in context-rich messaging environments. As a result, comparatively little work has examined how to identify online harassment in private conversations using methods that incorporate conversational and relational context. This gap motivates our first research question: how to identify online harassment in private messaging at scale using context-aware methods.

\subsection{Victim-Centered Harassment Interventions}

Research on how to respond to online harassment has identified several strategies available to victims, including seeking help, confronting the aggressor, ignoring or reframing the incident, seeking emotional or instrumental support, and using technical mechanisms such as reporting or blocking \cite{machackova2013effectiveness, o2023adolescent}. Other response strategies include factual correction, warning, empathy, humor, emotional bonding, and pointing out inconsistencies in harmful speech \cite{mathew2019thou, benesch2014countering}. However, the effectiveness of these strategies varies by context, severity, and relational dynamics; ignoring or disconnecting may help in some cases but can also backfire, while confrontational approaches may restore a sense of control but risk escalation \cite{varela2022ignore, black2010victim, craig2007responding}.

Existing platform interventions largely focus on reporting, blocking, moderation, or perpetrator-facing approaches such as counterspeech \cite{instagram-report,facebook_help_center, schieb2016governing, benesch2014countering}. Counterspeech research has examined how fact-based arguments, empathy, humor, warnings, and moral appeals may reduce harmful speech, but findings remain mixed \cite{mathew2019thou, hangartner2021empathy}. Prior work also cautions that interventions should be evaluated in real-world contexts, attend to behavioral and social consequences, and preserve transparency and user agency \cite{chung2023understanding, garland2022impact, reusser2021assessing, schieb2016governing}.

Psychological theories of coping, social support, and self-efficacy help explain why victim-centered interventions may be needed during harassment experiences. Coping theory emphasizes that individuals respond to stressful events through both problem-focused and emotion-focused strategies \cite{lazarus1984stress}. Social support theory suggests that support can buffer the negative effects of stressful experiences \cite{cohen1985stress}, while self-efficacy theory highlights the importance of perceived agency in responding to difficult situations \cite{bandura1977self}. These frameworks suggest that an effective intervention should not only address the perpetrator’s behavior, but also help the victim interpret the situation, regulate emotional burden, and feel more able to respond.

Recent AI-mediated systems point toward new possibilities for such support. \citet{chang2022thread} used synthetic responses to help users de-escalate contentious online discussions, while \citet{kim2024respect} developed systems for facilitating counter-harassment at scale. However, other work suggests that LLM-generated counterspeech may not always outperform simpler or non-contextualized interventions \cite{bar2024generative}. In addition, AI-mediated communication raises broader questions about how computational agents modify, augment, or generate messages on behalf of users \cite{hancock2020ai}. These questions are especially salient in private messaging, where AI-generated responses may operate as victim-facing support rather than public moderation.

Together, prior work leaves two key gaps. First, limited research has examined context-aware detection of harassment in private messaging. Second, although existing interventions often focus on moderation or perpetrator behavior, less work has investigated AI-generated responses designed to support victims during harassment in private conversations. These gaps motivate our study of LLM-based detection and response generation in adolescent Instagram DMs.

\section{Data}

\paragraph{Participant recruitment and study context.} The data used in this study originated from a larger longitudinal research effort designed to investigate social media signals associated with suicide risk and adverse mental health outcomes among youth. Participants involved adolescents with histories of suicide attempts, suicidal ideation, cybervictimization, and other mental health risk factors, as well as non-psychiatric controls. Specifically, the study included youth who had experienced recent suicide-related crises, sexual and gender minority youth with experiences of cyberbullying, and comparison participants without histories of suicidality or severe psychopathology. Participants were required to be English-speaking, to have at least one social media account, and to provide informed consent or assent for retrospective social media data donation.

Following consent procedures, participants exported and donated their social media archives through a secure web-based portal developed by the research team. The donated archives included retrospective social media activity such as posts, interactions, and direct (private) messages exchanged on social media platforms, including Instagram. Prior to analysis, the data underwent automated de-identification procedures designed to remove personally identifiable information through natural language processing and machine learning techniques. Specifically, software-based de-identification methods were used to remove or anonymize names, locations, and other potentially identifying textual content before researchers accessed the data. These procedures were designed to preserve the usability of conversational data while minimizing privacy risks associated with sensitive social media archives. The de-identified data were stored using secure and HIPAA-compliant infrastructure.

\paragraph{Instagram direct message dataset.} For the present study, we focused specifically on Instagram direct message conversations because they provided rich conversational context for studying online harassment in private messaging environments. The final dataset consisted of Instagram direct message conversations donated by 26 adolescents between the ages of 12 and 18, including some youth with suicide risk factors. The average age of the donors was 15.67 years, with 68\% being female donors and 32\% identifying as male. 76\% of the participants were white, 20\% Black, and 4\% of multiple races. 8\% identified as Hispanic. In total, the donated archives contained 80,053 Instagram direct messages. Because the data were collected from populations at elevated risk for suicidality and cybervictimization, the dataset provided a unique opportunity to study online harassment in a clinically and socially vulnerable population that may be especially in need of timely support and intervention.

\begin{figure*}[t]
    \centering
    \includegraphics[width=.85\linewidth]{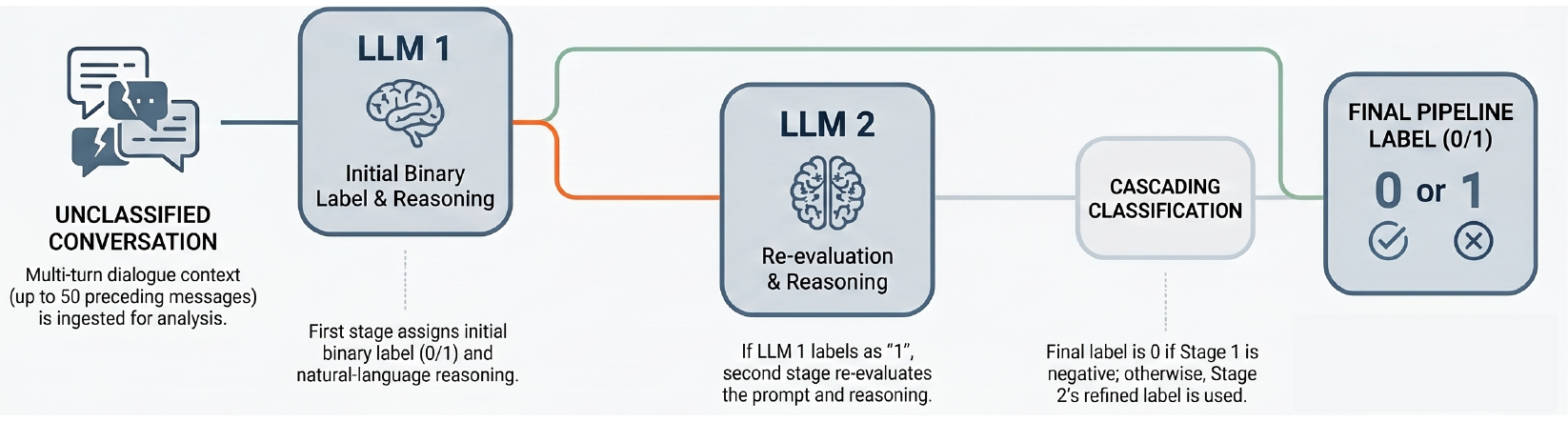}
    \caption{LLM classification pipeline.}
    \label{fig:class_pipeline_structure}
    \vspace{-.15in}
\end{figure*}

\section{RQ1: Detecting Online Harassment in Private Message Conversations}
\subsection{Human Labeling of Online Harassment}

\paragraph{Annotation task and labeling framework.} To construct ground-truth labels for detecting online harassment in Instagram DM conversations, human annotators reviewed messages within their surrounding conversational context and determined whether each target message constituted online harassment. For each target message, annotators were shown the preceding 10 messages in the conversation to provide sufficient interpersonal and conversational context for interpretation. We adopted the definition of online harassment as ``\textit{interpersonal aggression or offensive behavior(s) that is communicated over the internet or through other electronic media}'' \cite{slaughter2022new}. Labelers were instructed to determine whether the target message contained abusive, threatening, degrading, hostile, or otherwise harmful behavior directed toward an individual or group. The annotation task therefore emphasized contextual interpretation rather than isolated message classification, as many messages could only be interpreted accurately in relation to preceding conversational turns, sarcasm, escalation patterns, or ongoing interpersonal conflict. Each message received a binary label indicating whether it represented online harassment (1) or not (0).

\paragraph{Human annotators and labeling guidelines.} A total of seven human labelers participated in the annotation process. Labelers were provided with detailed written instructions that included definitions of online harassment, annotation guidelines, examples of labeled messages, and reminders regarding secure data handling practices. They were also instructed to avoid over-interpreting ambiguous language, emojis, slang, or interpersonal disagreement without sufficient contextual evidence, and to focus specifically on targeted harmful behavior. All labelers were young adults who had some form of prior background or personal experience related to mental health, online harassment, or social media.

\paragraph{Annotation reliability and adjudication.} To improve annotation reliability, each message was independently labeled by two annotators, with the second annotator completing the task without access to labels assigned during the first round. Across all doubly-labeled messages, 97.2\% received the same label from both annotators. Despite this high percentage agreement, the inter-rater agreement metric (Cohen's kappa) was relatively low ($\kappa = 0.112$), likely due to the highly imbalanced distribution of online harassment labels and the contextual subjectivity of the task. Therefore, messages with disagreements between the first two annotators were reviewed by a third labeler, who resolved conflicts by assigning the final
label, resulting in complete consensus across the dataset. This annotation process produced a total of 14,604 labeled Instagram messages.



We then excluded 7,076 messages sent by the data donors themselves, as our goal was to evaluate online harassment directed toward the participants and to subsequently assess the helpfulness of AI-generated responses for those individuals. After filtering these messages, the final dataset consisted of 7,528 labeled messages, including 41 messages labeled as online harassment. Three messages used as prompt examples in the classification pipeline were removed from evaluation, resulting in a final set of 7,525 messages used to evaluate the LLM classification pipeline (RQ1) and portions of the AI-generated response framework (RQ2).

\subsection{Classification Approach}

\paragraph{Motivation for context-aware classification.} Our human labeling process revealed that identifying online harassment in private messaging environments is highly context dependent. Many messages could not be accurately classified in isolation, as their meaning often depended on prior conversational exchanges, sarcasm, escalation patterns, and interpersonal dynamics between participants. These observations motivated the use of large language models (LLMs), which are well suited for reasoning over long conversational contexts and interpreting nuanced natural language interactions. Unlike traditional classifiers that typically operate on
individual posts or short text segments, LLMs can leverage multi-turn conversational histories to infer intent, relational context, and the severity of harmful interactions.

\paragraph{Motivation for cascading classification.} However, during early experimentation, we observed that a single-
stage LLM classifier was often overly sensitive and prone to misclassifications, particularly in ambiguous conversational settings involving slang, joking, or emotionally charged exchanges. To address this issue, we developed a two-stage cascading LLM pipeline, illustrated in
\Cref{fig:class_pipeline_structure}. 

\paragraph{Pipeline architecture and prompting framework.} This LLM-based classification pipeline was designed to explicitly incorporate conversational context when detecting online harassment in private messaging environments. We implemented the classifier using vLLM and the \textit{meta-llama/Llama-4-Scout-17B-16E-Instruct} model. For each target message, the model was provided with the preceding 50 messages from the conversation as contextual input. Messages were formatted sequentially as a dialogue, with each message presented on a separate line, enabling the model to reason over longer interpersonal and conversational histories. Similar to the human annotation process, the LLMs were instructed to classify only the final message in the conversation while using the preceding conversational turns as supporting context.

The classification pipeline combined the outputs of two sequential LLM stages, LLM 1 and LLM 2. For each classification task, LLM 1 received a system prompt that defined the classification rules and online harassment guidelines (Table~\ref{tab:a1_system_prompt}), together with a user prompt containing the conversation to be classified (Table~\ref{tab:a1_user_prompt}). The output generated by LLM 1 was subsequently passed to LLM 2, which also received its own system and user prompts (Table~\ref{tab:a2_system_prompt} and Table~\ref{tab:a2_user_prompt} respectively) designed to reevaluate the classification decision more conservatively. The system and user prompts included the definition of online harassment, annotation guidelines, and illustrative examples. We iteratively refined these prompts over dozens of prompt-engineering cycles to improve performance and calibrate the sensitivity of the model to online harassment.



\paragraph{Reasoning-guided cascading classification.} Both LLMs were instructed to generate structured outputs
consisting of (1) a binary classification label and (2) a brief natural-language explanation of the reasoning underlying the classification decision. The output generated by the first LLM, including both its label and reasoning, was subsequently provided as part of the input to the second LLM. This design allowed the second-stage model to reconsider the initial prediction while incorporating both the conversational context and the prior model's rationale. The second-stage prompts were specifically designed to reduce false positive classifications and encourage more conservative decision making in ambiguous cases. Requiring reasoning outputs
also facilitated qualitative error analysis and informed subsequent prompt refinement during development.

\paragraph{Final classification decision process.} We adopted a cascading decision framework for final classification. If the first-stage LLM classified a message as non-harassing (0), the final output was automatically assigned as non-harassing. Only messages classified as harassing (1) by the first-stage model were passed to the second-stage LLM for further evaluation, with the second-stage prediction becoming the final output. This design allowed the pipeline to
retain sensitivity to potentially harmful content while reducing unnecessary false positives. Final predictions produced through the cascading framework were compared against the human-labeled ground truth to generate classification reports and confusion matrices. Prompt iterations and pipeline refinements were guided by classification performance on the
labeled evaluation set.

\begin{table}[t]
\centering
\footnotesize
\setlength{\tabcolsep}{5pt}
\renewcommand{\arraystretch}{1.15}
\begin{tabular}{lcccc}
\toprule
\textbf{Class} & \textbf{Precision} & \textbf{Recall} & \textbf{F1-score} & \textbf{Support} \\
\midrule
0 & 0.9978 & 0.9854 & 0.9916 & 7485 \\
1 & 0.1805 & 0.6000 & 0.2775 & 40 \\
\midrule
\textbf{Accuracy} & \multicolumn{3}{c}{0.9834} & 7525 \\
\textbf{Macro avg} & 0.5891 & 0.7927 & 0.6345 & 7525 \\
\textbf{Weighted avg} & 0.9935 & 0.9834 & 0.9878 & 7525 \\
\bottomrule
\end{tabular}
\caption{Classification results given by the LLM pipeline, on the human-labeled online harassment dataset.} \label{tab:classification_report}
\vspace{-.1in}
\end{table}

\begin{table}[t]
\centering
\footnotesize
\setlength{\tabcolsep}{4pt}
\renewcommand{\arraystretch}{1.1}

\begin{minipage}{0.32\columnwidth}
\centering
(a) LLM pipeline

\vspace{2pt}

\begin{tabular}{lcc}
\toprule
 & \textbf{0} & \textbf{1} \\
\midrule
\textbf{0} & 7376 & 109 \\
\textbf{1} & 16 & 24 \\
\bottomrule
\end{tabular}
\end{minipage}
\hfill
\begin{minipage}{0.32\columnwidth}
\centering
(b) toxic-bert

\vspace{2pt}

\begin{tabular}{lcc}
\toprule
 & \textbf{0} & \textbf{1} \\
\midrule
\textbf{0} & 7350 & 135 \\
\textbf{1} & 27 & 13 \\
\bottomrule
\end{tabular}
\end{minipage}
\hfill
\begin{minipage}{0.32\columnwidth}
\centering
(c) Ensemble

\vspace{2pt}

\begin{tabular}{lcc}
\toprule
 & \textbf{0} & \textbf{1} \\
\midrule
\textbf{0} & 7403 & 82 \\
\textbf{1} & 24 & 16 \\
\bottomrule
\end{tabular}
\end{minipage}

\caption{Confusion matrices for the LLM pipeline and baseline classifiers on the human-labeled harassment dataset.}
\label{tab:confusion_matrices_all}
\vspace{-.1in}
\end{table}


\begin{table*}[t!]
\centering
\small
\begin{tabular}{p{0.31\linewidth} p{0.31\linewidth} p{0.31\linewidth}}
\toprule
\textbf{True Positive Example 1} & \textbf{True Positive Example 2} & \textbf{True Positive Example 3} \\
\midrule

\textbf{Person 2:} [NAME]? 

\textbf{Person 2:} I get B+ from the course 

\textbf{Person 2 (labeled message):} however I simply want that slut perish

&
\textbf{Person 2:} she’s fake platstic lol 

\textbf{Person 1:} Her dress makes her look better. she seems ordinary without that 

\textbf{Person 2 (labeled message):} Well she’s a idiot of fake platstic

&
\textbf{Person 2:} what can? 

\textbf{Person 2:} humiliating 

\textbf{Person 1 (labeled message):} Your entire face look so humiliating

\\

\bottomrule
\end{tabular}
\caption{Examples of true positive online harassment classifications by the LLM pipeline. In all three cases, the ground truth label and the LLM pipeline prediction were both positive (1).}
\label{tab:true_positive_examples}
\vspace{-.1in}
\end{table*}

\subsection{RQ1 Results}

\subsubsection{Performance of the LLM pipeline classifier}

We evaluated the performance of the proposed LLM classification pipeline against the human-labeled ground truth dataset consisting of 7,525 Instagram messages. Three messages used as prompt examples during development were excluded from evaluation. The resulting classification report and confusion matrix are presented in Table~\ref{tab:classification_report} and Table~\ref{tab:confusion_matrices_all}(a), respectively. Overall, the LLM pipeline achieved strong classification performance despite the highly imbalanced nature of the dataset, in which only 40 messages were labeled as online harassment. The classifier achieved an overall accuracy of 0.98 and a macro F1-score of 0.63. Importantly, for the positive class corresponding to online harassment, the pipeline achieved a recall of 0.60 and an F1-score of 0.28. These results indicate that the classifier was able to identify a substantial proportion of harassment messages while maintaining high overall precision on non-harassing messages.

The confusion matrix further illustrates the behavior of the pipeline in practice. Out of 40 human-labeled online harassment messages, the classifier correctly identified 24 cases, while missing 16 cases. At the same time, the pipeline produced 109 false positives across 7,485 non-harassing messages. Given the contextual ambiguity and conversational subjectivity inherent in private messaging environments, we prioritized sensitivity to harmful content over aggressively minimizing false positives. This design choice was motivated by the downstream intervention-oriented goals of the system (RQ2), where failing to identify harmful content may carry greater consequences than over-flagging potentially harmful messages for further review.

The cascading two-stage architecture substantially improved classification behavior compared to the initial single-stage LLM approach; see examples in Table~\ref{tab:true_positive_examples}. 
Specifically, false positives decreased from 156 after the first-stage LLM to 109 following cascading classification, while preserving the pipeline’s ability to detect harmful content. These findings suggest that multi-stage LLM architectures may offer a practical mechanism for balancing sensitivity and specificity in context-dependent online harm detection tasks.

\subsubsection{Error analysis of false negatives}

To better understand the limitations of the proposed classification pipeline, we qualitatively examined several false negative cases in which the ground truth label was online harassment but the pipeline classified the message as non-harassing. Across these cases, we observed that the pipeline was more likely to miss harassment when the harmful language was indirect, conversationally embedded, or lacked explicit targeting cues. For example, one paraphrased false negative message stated, ``\textit{Those people are really fucking idiotic},'' while another stated ``\textit{The entirety of her pictures are really hideous}.'' Although both messages contain derogatory language, the harassment target is conversationally implied rather than directly specified. Similarly, a message like ``\textit{I've already loathed him for a long time, like many years}'' expressed sustained hostility toward another individual but lacked explicit abusive phrasing.

These examples suggest that the conservative second-stage filtering process, while effective for reducing false positives, may also suppress some contextually harmful but less explicit forms of harassment. Broadly, the false negative cases highlight the difficulty of distinguishing between general negativity, interpersonal conflict, and targeted online harassment in conversational settings. Since our downstream response framework prioritized sensitivity to harmful interactions, improving recall for indirect and relational forms of harassment remains an important direction for future work.

\subsubsection{Comparison with baseline classifiers}

To contextualize the performance of our proposed pipeline, we compared it against two baseline approaches representing commonly used paradigms for harmful language detection: (1) a widely used pretrained toxicity classifier and (2) a traditional supervised ensemble learning framework.

First, we evaluated the pretrained \textit{unitary/toxic-bert} model \cite{Detoxify}, which has been broadly used for toxicity and abusive language detection in online text. We used toxic output head on the same human-labeled dataset, excluding the prompt examples used during pipeline development. We experimented with providing both entire conversations and isolated messages as inputs. However, the model performed substantially better on individual messages, likely because it was pretrained primarily on short-form toxicity classification tasks rather than longer multi-turn conversations. The final threshold for classification was optimized using F1-score over a fixed grid search.

Second, we constructed a supervised ensemble baseline following prior work on cyberbullying and online harassment detection \cite{azeez2023classification}. The ensemble was trained using existing publicly available datasets containing online harassment and cyberbullying labels from multiple social media platforms \cite{shahane_2020, golbeck2017large, hamlett2022labeled}. After preprocessing, oversampling, and train-test splitting, we trained and fine-tuned 30 classifiers spanning multiple model families, including Naive Bayes, Logistic Regression, Support Vector Machines, XGBoost, and DistilBERT \cite{sanh2019distilbert}. Final predictions were generated through majority voting.

Tables~\ref{tab:confusion_matrices_all} and~\ref{tab:baseline_comparison} present the macro-level performance comparison across all three classifiers; Table~\ref{tab:baseline_classification_reports} gives additional details about the performance of the two baselines. Overall, our proposed LLM pipeline outperformed both baselines, particularly on the online harassment class. The \textit{toxic-bert} baseline achieved a positive-class F1-score of only 0.14, substantially lower than the LLM pipeline’s 0.28. Similarly, the ensemble achieved a positive-class F1 of 0.23, also below the proposed approach. The comparatively weak performance of the  toxicity model suggests that classifiers trained primarily on isolated public-facing toxic content may not generalize to the conversational and context-dependent nature of private messaging harassment. Likewise, although the ensemble baseline incorporated multiple architectures and training datasets, its lower recall and F1-score indicate the difficulty of transferring publicly trained harassment classifiers to private conversational environments. 

These findings demonstrate the importance of conversational context in detecting  harassment in private messaging. The superior performance of our LLM pipeline suggests that context-aware LLM approaches may be better suited for identifying nuanced, relational, and multi-turn forms of harm that are difficult to capture using conventional toxicity classifiers or traditional supervised learning methods.

\begin{table}[t]
\centering
\footnotesize
\setlength{\tabcolsep}{6pt}
\renewcommand{\arraystretch}{1.2}
\begin{tabular}{lccccc}
\toprule
Classifier & Precision & Recall & F1 & Positive F1\\
\midrule
LLM pipeline & \textbf{0.5891} & \textbf{0.7927} & \textbf{0.6345} & \textbf{0.2775}\\
toxic-bert & 0.5421 & 0.6535 & 0.5637 & 0.1383\\
Ensemble & 0.5800 & 0.6945 & 0.6124 & 0.2319\\
\bottomrule
\end{tabular}
\caption{Macro performance comparison of LLM pipeline classifier with baselines.} \label{tab:baseline_comparison}
\vspace{-.15in}
\end{table}

\section{RQ2: AI-generated responses}

Building on RQ1, which focused on identifying online harassment in private messaging environments, RQ2 investigates whether LLMs can help support individuals experiencing online harassment through context-aware response generation. 
LLMs offer the ability to generate personalized, context-sensitive, and interactive language in real time. We therefore explore whether LLMs can generate responses that are perceived as emotionally supportive, de-escalatory, and practically helpful to tackle private online harassment.

\subsection{Approach to Response Generation}



\begin{figure}
    \centering
    \includegraphics[width=1\linewidth]{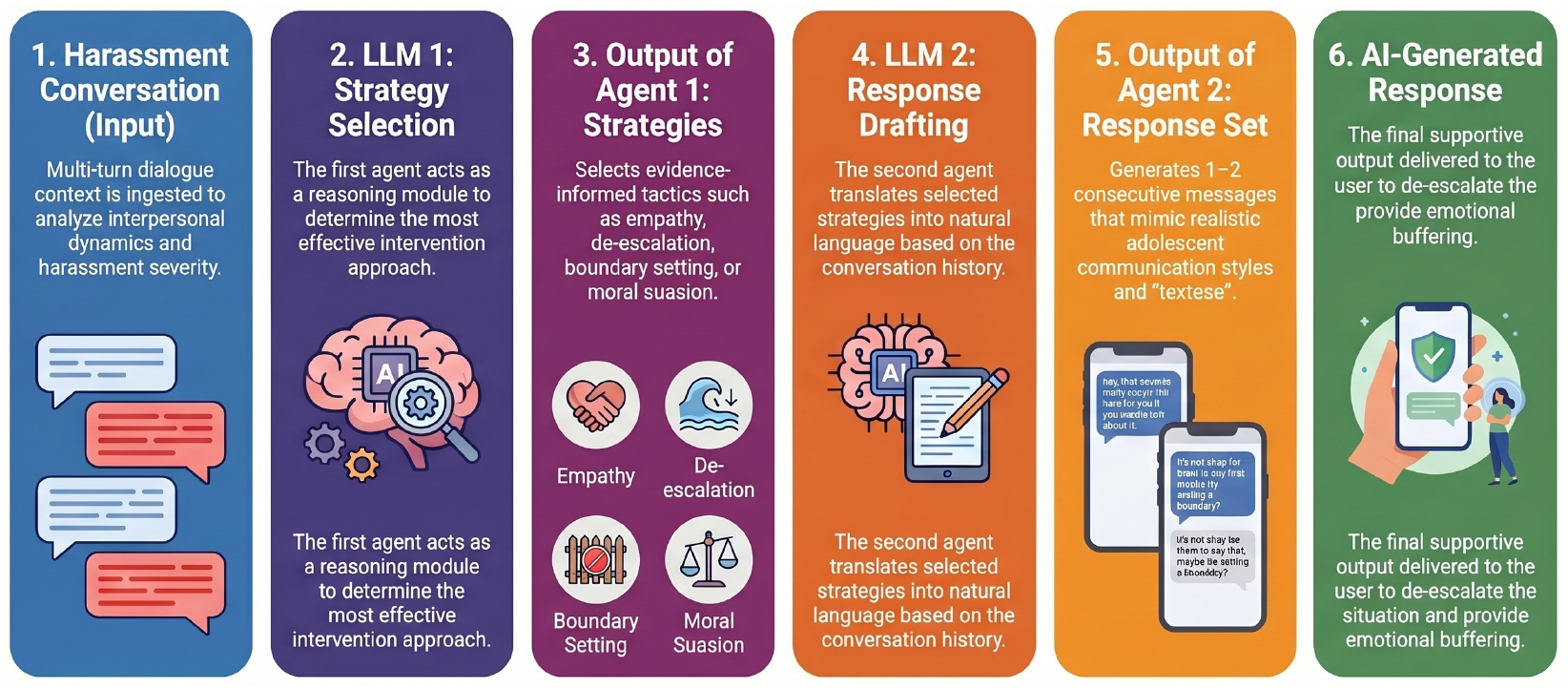}
    \caption{Response generation pipeline.}
    \label{fig:SR_structure}
    \vspace{-.1in}
\end{figure}


\paragraph{Motivation for context-aware response generation.} Our response generation framework is motivated by three core considerations. First, online harassment in private messaging environments is highly contextual and relational, requiring an understanding of conversational history, tone, and interpersonal dynamics. Second, effective responses to online harassment are often strategic rather than purely reactive, involving mechanisms such as empathy, de-escalation, boundary setting, or moral reframing. Third, because adolescents frequently communicate through informal and platform-specific language styles, generated responses must balance helpfulness with conversational naturalness.

\paragraph{Response generation pipeline architecture.} To address these challenges, we designed a two-stage LLM response generation pipeline illustrated in Figure~\ref{fig:SR_structure}. The pipeline used the same \textit{meta-llama/Llama-4-Scout-17B-16E-Instruct} model as in RQ1. The input to the pipeline consisted of conversation contexts containing online harassment messages identified either through human labeling or through the LLM classification pipeline developed for RQ1. By leveraging both human-labeled and pipeline-detected harassment messages, we constructed a broader dataset of conversational scenarios for response generation.

\paragraph{Strategy-guided response generation.} Instead of generating responses in a single step, we adopted a staged pipeline architecture to separate strategic reasoning from linguistic realization. During early experimentation, we observed that direct one-step response generation often produced generic, inconsistent, or insufficiently supportive outputs. We therefore designed the first-stage LLM (LLM 1) to function as a strategy selection module. Given the conversation context, LLM 1 selected one or more response strategies and generated a brief rationale for those choices. This stage grounded response generation in prior literature on coping strategies, counterspeech, de-escalation, and online conflict mitigation \cite{mathew2019thou, hangartner2021empathy, munger2021don, young2024responding, reusser2021assessing}. Specifically, we incorporated seven evidence-informed response strategies, including warning the harasser of consequences, denouncing hateful behavior, restoring affective connection, identifying hypocrisy, using empathy, applying moral suasion, and benevolently correcting harmful behavior.

The second-stage LLM (LLM 2) then generated the actual response messages conditioned on both the original conversation context and the strategic guidance produced by LLM 1. This design allowed the response generation process to remain grounded in theoretically motivated intervention strategies while preserving contextual flexibility and conversational realism. The system and user prompts used for both LLM stages are provided in Tables \ref{tab:Simulated_response_Agent_1_system_prompt}, \ref{tab:Simulated_response_Agent_1_user_prompt}, \ref{tab:Simulated_response_Agent_2_system_prompt}, and \ref{tab:Simulated_response_Agent_2_user_prompt}. 

\paragraph{Adolescent language style and realism constraints.} To improve ecological validity, we constrained the generated responses to resemble realistic adolescent communication styles observed in the donated Instagram conversations. We first calculated the 25th and 75th percentile ranges of message length and number of consecutive responses across the original participant responses. Based on these distributions, we instructed LLM 2 to generate between one and two consecutive response messages, each containing approximately 3 to 13 words. In addition, because adolescent online communication frequently involves abbreviated and informal language styles, we instructed the model to optionally use textese, abbreviations, expressive lengthening, and reduced grammatical formality where contextually appropriate \cite{di2024patterns}. At the same time, prompts explicitly instructed the model to avoid retaliatory escalation and to prioritize de-escalation, emotional support, and conversational appropriateness.


\subsection{Evaluation Approach of Generated Responses}


\newcolumntype{N}[1]{>{\raggedleft\arraybackslash}p{#1}}
\newcolumntype{T}{>{\raggedright\arraybackslash}X}

\begin{table}[t]
\centering
\begingroup
\small
\setlength{\tabcolsep}{6pt}
\renewcommand{\arraystretch}{1.15}

\begin{tabularx}{\columnwidth}{
  >{\raggedleft\arraybackslash}p{0.1cm}
  >{\raggedright\arraybackslash}X
}
\hline
\textbf{\#} & \textbf{Question} \\
\hline
1 & Which response set is more effective in helping the user stop online harassment \cite{machackova2013effectiveness}?  \\
2 & Which response set is more effective in deescalating the situation \cite{machackova2013effectiveness}? \\
3 & Which response set puts the user in a better position against the harasser \cite{machackova2013effectiveness}? \\
4 & Which response set is more emotionally helpful to the user \cite{machackova2013effectiveness}? \\
5 & Which response set sounds more natural in the conversation? \\
6 & If no response set is ``Ignoring'', is ignoring the harasser a better option? \cite{machackova2013effectiveness}? \\
\hline
\end{tabularx}
\endgroup
\caption{AI-generated response evaluation rubric.}\label{tab:sr_evaluation_question_table}
\vspace{-.1in}
\end{table}

\paragraph{Motivation and principles of the evaluation approach.}

To evaluate the perceived effectiveness of the AI-generated responses, we designed a comparative human evaluation framework grounded in two core principles. First, because online harassment unfolds within rich interpersonal and conversational contexts, responses should be evaluated within the surrounding conversation rather than in isolation. Second, the effectiveness of responses to online harassment cannot be assessed solely based on linguistic quality; instead, evaluation must consider whether responses are perceived as emotionally supportive, de-escalatory, and helpful in addressing the harassment situation. Accordingly, our evaluation framework focused on comparing AI-generated responses against the original responses written by participants in the Instagram conversations, allowing us to assess whether the generated responses offered meaningful advantages in realistic online harassment scenarios.

\paragraph{Overview of the evaluation task.}

We first constructed a sample of 100 online harassment cases. Consecutive online harassment messages originating from the same non-donor user/participant were grouped into a single harassment case during sampling. Cases were drawn from both the human-labeled online harassment dataset and the online harassment messages identified by the LLM classification pipeline developed in RQ1. Human-labeled harassment cases were preferentially included unless overlap constraints required exclusion. The remaining cases were randomly sampled from the pipeline-classified positive cases, as the pipeline demonstrated stronger performance than the baseline classifiers.

For each harassment case, we extracted both the surrounding conversational context and the participant’s original response messages. To ensure readability and contextual relevance, the conversation context shown to annotators was capped at 21 messages and limited to messages occurring within 48 hours before the harassment event. This threshold helped to balance contextual completeness with cognitive burden for evaluators; notably, 21 messages captured the complete one-hour conversational history preceding the harassment event in approximately 75\% of cases.

Participant response messages were similarly constrained using temporal distance, interruption, and message-count criteria. Responses were limited to those occurring within 24 hours following the harassment event. To reduce redundancy across evaluation samples, we excluded overlapping conversational contexts and avoided cases where another harassment event occurred between the target harassment message and the participant's response. In situations where participants did not respond within a reasonable conversational or temporal window, ``Ignoring'' was recorded as the original human response condition.

Using these materials, we created a blinded comparative labeling task in which the participant’s original responses and the AI-generated responses were randomly assigned to two unlabeled columns. Annotators therefore evaluated the responses without knowing which set was AI-generated.

The evaluation task was completed independently by 3 individuals. All evaluators were young adults with prior background or lived experience related to mental health, online harassment, or social media, and one evaluator also participated in the earlier online harassment labeling task. 

\begin{table}[t]
\centering
\footnotesize
\setlength{\tabcolsep}{2pt}
\renewcommand{\arraystretch}{1.2}
\begin{tabular}{lc | lc}
\toprule
 & $\kappa$ after adjudication & & $\kappa$ after adjudication  \\
\midrule
Question 1 & 1.000000  & Question 4 & 0.982980  \\
Question 2 & 0.965566 & Question 5 & 1.000000  \\
Question 3 & 0.983342 & Question 6 & 0.970732  \\
\bottomrule
\end{tabular}
\caption{Two-way Fleiss' $\kappa$ in response set comparison.}
\label{tab:fleiss_kappa}
\vspace{-.1in}
\end{table}

\paragraph{Overview of evaluation questions and rubric design.}

For each of the 100 cases, annotators answered the six evaluation questions shown in Table~\ref{tab:sr_evaluation_question_table}. The evaluation rubric was designed to assess two primary dimensions: (1) perceived helpfulness and (2) conversational naturalness.

Questions 1-4 were grounded in prior work on coping strategies for cyberbullying victims \cite{machackova2013effectiveness}. Specifically, these questions evaluated whether a response was perceived as effective in stopping harassment, de-escalating the interaction, improving the victim’s position relative to the harasser, and providing emotional support. Together, these measures operationalized the broader construct of response helpfulness. Question 5 evaluated whether the response sounded natural within the context of the original conversation, while Question 6 assessed whether ignoring the harassment would have been preferable to responding at all.

For Questions 1-5, the human evaluators selected among four response options: ``Response set 1,'' ``Response set 2,'' ``No preference,'' or ``Both response sets make things worse.'' These options were designed to capture a broad range of comparative judgments while avoiding forced preferences. Question 6 used a binary ``Yes'' or ``No'' response format.

\subsection{RQ2 Results}

All three evaluators completed the comparative evaluation task for the 100 online harassment cases. Because the evaluation involved nuanced judgments regarding emotional support, de-escalation, and conversational appropriateness, disagreements among evaluators were expected and often reflected differing interpretations of context and severity. To improve consistency and establish shared interpretations of the evaluation rubric, we conducted four adjudication meetings during which evaluators discussed disagreements and clarified ambiguous cases. 
Following adjudication, the average Fleiss’ Kappa across Questions 1-5 increased substantially from 0.28 to 0.98, indicating near-complete agreement among evaluators. Final inter-rater agreement values for each question are presented in Table~\ref{tab:fleiss_kappa}.

Across Questions 1-4, which collectively operationalized response helpfulness, evaluators showed a strong overall preference for the AI-generated responses (95\% CI: 0.767--0.815, $p < .001$). Excluding two tied cases, AI-generated responses were preferred in 80.2\% of the evaluated harassment scenarios. These findings suggest that the generated responses were consistently perceived as more emotionally supportive, de-escalatory, and practically helpful than the original participant responses.

In contrast, evaluators preferred the original participant responses in terms of conversational naturalness (Question 5). The confidence interval associated with preference for AI-generated responses on naturalness was substantially lower (95\% CI: 0.035--0.089, $p < .001$), suggesting that although the generated responses were often perceived as more helpful, they did not consistently match the linguistic authenticity or conversational style of naturally occurring adolescent responses.

For Question 6, evaluators generally did not consider ignoring the harassment to be preferable to responding (95\% CI: 0.125--0.218, $p < .001$). Across applicable cases, evaluators indicated that ignoring would have been preferable in only 17.1\% of cases. However, post-adjudication discussions revealed that evaluators viewed the severity of harassment as an important moderating factor. Less severe or joking forms of harassment were sometimes perceived as situations where ignoring may be more appropriate, whereas more severe harassment scenarios motivated stronger preferences for active responses. Overall, regardless of whether ignoring was viewed as preferable in specific situations, AI-generated responses were consistently perceived as more helpful than the original participant responses.

\section{Discussion}

Our findings demonstrate the potential of large language models to support both the detection of and response to online harassment in private messaging environments. In RQ1, we showed that conversational context is critical for identifying online harassment in private conversations, and that an LLM-based cascading pipeline can outperform conventional toxicity classifiers trained primarily on public social media data. In RQ2, we found that AI-generated responses grounded in conversational context and evidence-based intervention strategies were consistently perceived as more helpful than the original participant responses. Together, these findings point toward the potential of context-aware and victim-centered AI support systems for addressing online harassment among vulnerable youth.

\subsection{Context-Aware Harassment Detection Implications}

Our findings highlight the importance of conversational context in detecting online harassment within private messaging environments. Unlike public-facing toxicity classification tasks that often rely on isolated posts, harassment in private conversations frequently unfolds across multiple conversational turns and depends on interpersonal dynamics, sarcasm, escalation, and relational history. This distinction was reflected both in our human labeling process and in the stronger performance of the context-aware LLM pipeline relative to baseline classifiers trained primarily on public social media data.

More broadly, our results suggest that LLM-based approaches may be particularly useful for harmful-content detection tasks involving long conversational contexts and limited labeled data. In addition, the pipeline-based approach may help reduce annotator exposure to harmful content by partially automating large-scale labeling workflows while preserving human oversight. However, the modest performance on the positive class also underscores the need for careful evaluation and human validation before deploying such systems in real-world moderation settings.

\subsection{Implications for Victim-Centered Intervention}

Our findings suggest the potential of AI-generated responses as a form of victim-centered support during online harassment. 
Specifically, 
the strong evaluator preference for AI-generated responses in terms of perceived helpfulness suggests that responses grounded in conversational context and evidence-informed intervention strategies may provide more emotionally supportive and de-escalatory communication than spontaneous victim responses alone.

One mechanism that may help explain these findings is what we describe as \textit{synthetic buffering}: the use of AI-generated responses to partially absorb the emotional and cognitive burden associated with responding to online harassment. Prior work has discussed the role of AI-mediated communication in interpersonal interaction \cite{hancock2020ai}; our findings extend this idea toward a form of AI mediation designed specifically for harm reduction and de-escalation. In this framing, the AI system acts as an intermediary layer between the victim and the harassment interaction by interpreting the conversational context, identifying potentially appropriate response strategies, and drafting responses that the victim may adapt or use. Rather than requiring individuals to immediately formulate emotionally regulated responses while experiencing distress, frustration, or fear, the system may help externalize some of the cognitive and emotional labor associated with navigating harassment interactions.

This buffering mechanism may be especially relevant in private messaging environments, where harassment often unfolds rapidly and relationally across multiple conversational turns and bystanders may not be available to offer support~\cite{blackwell2018online}. By generating context-sensitive and strategy-informed responses, AI-mediated systems may help victims maintain conversational control, reduce escalation, and respond in ways that feel more emotionally supported or empowered. Importantly, this form of mediation differs from traditional moderation systems because it is directed toward assisting the victim rather than restricting the perpetrator. At the same time, the AI-generated responses remain user-facing and optional, preserving the user’s agency to decide whether and how to engage.

At the same time, our findings also reveal an important tension between helpfulness and conversational naturalness. Although evaluators consistently preferred AI-generated responses in terms of supportiveness and de-escalation, original participant responses were perceived as more natural within the conversational context. This suggests that intervention-oriented responses that are strategically effective may not always align with the linguistic authenticity of naturally occurring adolescent communication. Designing AI-generated responses that balance effectiveness, realism, and user agency therefore remains an important challenge for future AI-mediated support systems.

\subsection{Ethical Risks and Design Considerations}

While our findings point toward the potential of AI-mediated support systems for online harassment, they also raise important design and safety considerations. First, AI-generated responses should not be interpreted as replacements for human social support, professional intervention, or existing safety infrastructures. Over-reliance on automated responses may unintentionally reshape how individuals navigate interpersonal conflict online or reduce engagement with human sources of support. The long-term psychological and behavioral effects of repeated AI-mediated interactions therefore remain unclear.

Second, the use of AI-generated responses introduces important tensions around mediation and user agency. Unlike traditional human mediation, the ``synthetic buffering'' approach explored in this work is visible only to the victim and not to the perpetrator. This may help preserve user autonomy and privacy while enabling rapid support during moments of distress. However, covert forms of AI mediation may also alter conversational norms or influence interpersonal dynamics in ways that are difficult to anticipate. Preserving user control over whether, when, and how AI-generated responses are used is therefore critical.

Finally, these systems may also introduce risks of misuse or unintended escalation. AI-generated responses designed to support victims could potentially be repurposed to facilitate harassment, manipulation, or antagonistic interactions. In addition, although our prompts explicitly prioritized de-escalation and non-retaliatory behavior, generated responses may still produce undesirable outcomes in real-world interactions. Future deployment of such systems should therefore incorporate safeguards, human oversight, and careful evaluation of both intended and unintended consequences.

\subsection{Limitations and Future Work}

We recognize that this study has some limitations that motivate future research. First, the dataset consisted of Instagram direct messages donated by 26 adolescents, including youth with suicide risk factors, and therefore reflects a specific population and communication context. The relatively small number of online harassment cases also limits the generalizability of the classification results. Future work should evaluate these methods on larger and more diverse datasets across platforms and populations.

Second, our evaluation focused on perceived helpfulness rather than real-world outcomes. Although evaluators preferred the AI-generated responses across multiple dimensions of supportiveness, we do not know whether these responses would actually reduce harassment or improve well-being in practice. Future studies should therefore investigate the real-world psychological and behavioral impacts of AI-mediated responses.

Finally, both LLM pipelines remain imperfect. The classification system still produced false positives and false negatives, while the generated responses were perceived as less natural than original participant responses. Future work may explore improved models and personalization approaches to better balance helpfulness, realism, and safety.

\section{Conclusion}

In this paper, we investigated the use of large language models for detecting and responding to online harassment in private messaging environments. We developed a context-aware LLM classification pipeline that outperformed baseline classifiers on private Instagram conversations and introduced an LLM-based response generation framework grounded in prior intervention strategies. We further found that the AI-generated responses were consistently perceived as more helpful than the original participant responses. Together, these findings highlight the potential of context-aware and victim-centered AI systems to support vulnerable youth during experiences of online harassment.


{\fontsize{10pt}{9.5pt} \selectfont{\bibliography{aaai25}}

@article{suler2004online,
  title={The online disinhibition effect},
  author={Suler, John},
  journal={Cyberpsychology \& behavior},
  volume={7},
  number={3},
  pages={321--326},
  year={2004},
  publisher={Mary Ann Liebert, Inc.}
}

@book{lazarus1984stress,
  title={Stress, Appraisal, and Coping},
  author={Lazarus, Richard S. and Folkman, Susan},
  year={1984},
  publisher={Springer},
  address={New York}
}

@article{cohen1985stress,
  title={Stress, Social Support, and the Buffering Hypothesis},
  author={Cohen, Sheldon and Wills, Thomas A.},
  journal={Psychological Bulletin},
  volume={98},
  number={2},
  pages={310--357},
  year={1985},
  doi={10.1037/0033-2909.98.2.310}
}

@article{bandura1977self,
  title={Self-efficacy: Toward a Unifying Theory of Behavioral Change},
  author={Bandura, Albert},
  journal={Psychological Review},
  volume={84},
  number={2},
  pages={191--215},
  year={1977},
  doi={10.1037/0033-295X.84.2.191}
}

@inproceedings{alsoubai2025timeliness,
  title={Timeliness matters: leveraging reinforcement learning on social media data to prioritize high-risk conversations for promoting youth online safety},
  author={Alsoubai, Ashwaq and Park, Jinkyung Katie and Stringhini, Gianluca and Ma, Meiyi and De Choudhury, Munmun and Wisniewski, Pamela J},
  booktitle={Proceedings of the International AAAI Conference on Web and Social Media},
  volume={19},
  pages={37--51},
  year={2025}
}

@inproceedings{blackwell2018online,
  title={When online harassment is perceived as justified},
  author={Blackwell, Lindsay and Chen, Tianying and Schoenebeck, Sarita and Lampe, Cliff},
  booktitle={Proceedings of the International AAAI Conference on Web and Social Media},
  volume={12},
  number={1},
  year={2018}
}

@article{o2023adolescent,
  title={Adolescent help-seeking: an exploration of associations with perceived cause of emotional distress},
  author={O'Neill, Alisha and Stapley, Emily and Rehman, Ishba and Humphrey, Neil},
  journal={Frontiers in Public Health},
  volume={11},
  pages={1183092},
  year={2023},
  publisher={Frontiers Media SA}
}

@book{vogels2021state,
  title={The state of online harassment},
  author={Vogels, Emily A},
  volume={13},
  year={2021},
  publisher={Pew Research Center Washington, DC}
}

@article{black2010victim,
  title={Victim strategies to stop bullying},
  author={Black, Sally and Weinles, Dan and Washington, Ericka},
  journal={Youth violence and juvenile justice},
  volume={8},
  number={2},
  pages={138--147},
  year={2010},
  publisher={SAGE Publications Sage CA: Los Angeles, CA}
}

@article{bretschneider2014detecting,
  title={Detecting online harassment in social networks},
  author={Bretschneider, Uwe and W{\"o}hner, Thomas and Peters, Ralf},
  year={2014}
}

@article{yin2009detection,
  title={Detection of harassment on web 2.0},
  author={Yin, Dawei and Xue, Zhenzhen and Hong, Liangjie and Davison, Brian D and Kontostathis, April and Edwards, Lynne and others},
  journal={Proceedings of the Content Analysis in the WEB},
  volume={2},
  number={0},
  pages={1--7},
  year={2009},
  publisher={Madrid, Spain}
}

@inproceedings{golbeck2017large,
  title={A large labeled corpus for online harassment research},
  author={Golbeck, Jennifer and Ashktorab, Zahra and Banjo, Rashad O and Berlinger, Alexandra and Bhagwan, Siddharth and Buntain, Cody and Cheakalos, Paul and Geller, Alicia A and Gnanasekaran, Rajesh Kumar and Gunasekaran, Raja Rajan and others},
  booktitle={Proceedings of the 2017 ACM on web science conference},
  pages={229--233},
  year={2017}
}

@inproceedings{umarova2024xenophobia,
  title={Xenophobia Meter: Defining and Measuring Online Sentiment toward Foreigners on Twitter},
  author={Umarova, Khonzoda and Okorafor, Oluchi and Lu, Pinxian and Shan, Sophia and Xu, Alex and Zhou, Ray and Otiono, Jennifer and Lyon, Beth and Leshed, Gilly},
  booktitle={Proceedings of the International AAAI Conference on Web and Social Media},
  volume={18},
  pages={1517--1530},
  year={2024}
}

@article{zaretsky2024generative,
  title={Generative artificial intelligence to transform inpatient discharge summaries to patient-friendly language and format},
  author={Zaretsky, Jonah and Kim, Jeong Min and Baskharoun, Samuel and Zhao, Yunan and Austrian, Jonathan and Aphinyanaphongs, Yindalon and Gupta, Ravi and Blecker, Saul B and Feldman, Jonah},
  journal={JAMA network open},
  volume={7},
  number={3},
  pages={e240357--e240357},
  year={2024},
  publisher={American Medical Association}
}

@inproceedings{lee2024large,
  title={Large language models produce responses perceived to be empathic},
  author={Lee, Yoon Kyung and Suh, Jina and Zhan, Hongli and Li, Junyi Jessy and Ong, Desmond C},
  booktitle={2024 12th International Conference on Affective Computing and Intelligent Interaction (ACII)},
  pages={63--71},
  year={2024},
  organization={IEEE}
}

@article{wang2024tutor,
  title={Tutor copilot: A human-ai approach for scaling real-time expertise},
  author={Wang, Rose E and Ribeiro, Ana T and Robinson, Carly D and Loeb, Susanna and Demszky, Dora},
  journal={arXiv preprint arXiv:2410.03017},
  year={2024}
}

@article{slaughter2022new,
  title={New frontiers: Moving beyond cyberbullying to define online harassment},
  author={Slaughter, Autumn and Newman, Elana},
  journal={Journal of Online Trust and Safety},
  volume={1},
  number={2},
  year={2022}
}

@inproceedings{katsaros2022reconsidering,
  title={Reconsidering tweets: Intervening during tweet creation decreases offensive content},
  author={Katsaros, Matthew and Yang, Kathy and Fratamico, Lauren},
  booktitle={Proceedings of the International AAAI Conference on Web and Social Media},
  volume={16},
  pages={477--487},
  year={2022}
}

@article{hangartner2021empathy,
  title={Empathy-based counterspeech can reduce racist hate speech in a social media field experiment},
  author={Hangartner, Dominik and Gennaro, Gloria and Alasiri, Sary and Bahrich, Nicholas and Bornhoft, Alexandra and Boucher, Joseph and Demirci, Buket Buse and Derksen, Laurenz and Hall, Aldo and Jochum, Matthias and others},
  journal={Proceedings of the National Academy of Sciences},
  volume={118},
  number={50},
  pages={e2116310118},
  year={2021},
  publisher={National Academy of Sciences}
}

@inproceedings{davidson2017automated,
  title={Automated hate speech detection and the problem of offensive language},
  author={Davidson, Thomas and Warmsley, Dana and Macy, Michael and Weber, Ingmar},
  booktitle={Proceedings of the international AAAI conference on web and social media},
  volume={11},
  number={1},
  pages={512--515},
  year={2017}
}

@inproceedings{mathew2019thou,
  title={Thou shalt not hate: Countering online hate speech},
  author={Mathew, Binny and Saha, Punyajoy and Tharad, Hardik and Rajgaria, Subham and Singhania, Prajwal and Maity, Suman Kalyan and Goyal, Pawan and Mukherjee, Animesh},
  booktitle={Proceedings of the international AAAI conference on web and social media},
  volume={13},
  pages={369--380},
  year={2019}
}

@article{munger2021don,
  title={Don’t@ Me: Experimentally reducing partisan incivility on Twitter},
  author={Munger, Kevin},
  journal={Journal of Experimental Political Science},
  volume={8},
  number={2},
  pages={102--116},
  year={2021},
  publisher={Cambridge University Press}
}

@article{reusser2021assessing,
  title={Assessing the Prevalence of Benevolence in Response to Online Toxicity on Reddit: A First Step},
  author={Reusser, Alison I Young and Veit, Kristian M and Gassin, Elizabeth A and Case, Jonathan P and Reusser, George M},
  year={2021},
  publisher={PubPub}
}

@article{kim2024respect,
  title={ReSPect: Enabling Active and Scalable Responses to Networked Online Harassment},
  author={Kim, Haesoo and Lee, Juhoon and Jang, Jeong-Woo and Kim, Juho},
  journal={Proceedings of the ACM on Human-Computer Interaction},
  volume={8},
  number={CSCW1},
  pages={1--30},
  year={2024},
  publisher={ACM New York, NY, USA}
}

@article{young2024responding,
  title={Responding to Online Toxicity: Which Strategies Make Others Feel Freer to Contribute, Believe That Toxicity Will Decrease, and Believe That Justice Has Been Restored?},
  author={Young Reusser, Alison I and Veit, Kristian M and Gassin, Elizabeth A and Case, Jonathan P},
  journal={Collabra: Psychology},
  volume={10},
  number={1},
  pages={92328},
  year={2024},
  publisher={University of California Press}
}

@article{chang2022thread,
  title={Thread with caution: Proactively helping users assess and deescalate tension in their online discussions},
  author={Chang, Jonathan P and Schluger, Charlotte and Danescu-Niculescu-Mizil, Cristian},
  journal={Proceedings of the ACM on human-computer interaction},
  volume={6},
  number={CSCW2},
  pages={1--37},
  year={2022},
  publisher={ACM New York, NY, USA}
}

@article{craig2007responding,
  title={Responding to bullying: What works?},
  author={Craig, Wendy and Pepler, Debra and Blais, Julie},
  journal={School psychology international},
  volume={28},
  number={4},
  pages={465--477},
  year={2007},
  publisher={Sage Publications Sage UK: London, England}
}

@article{machackova2013effectiveness,
  title={Effectiveness of coping strategies for victims of cyberbullying},
  author={Machackova, Hana and Cerna, Alena and Sevcikova, Anna and Dedkova, Lenka and Daneback, Kristian},
  journal={Cyberpsychology: Journal of Psychosocial Research on Cyberspace},
  volume={7},
  number={3},
  year={2013}
}

@article{article,
author = {Perren, Sonja and Corcoran, Lucie and Cowie, Helen and Dehue, Francine and Garcia, D'Jamila and Mc Guckin, Conor and Sevcikova, Anna and Tsatsou, Panayiota and Völlink, Trijntje},
year = {2012},
month = {01},
pages = {283-292},
title = {Tackling Cyberbullying: Review of Empirical Evidence Regarding Successful Responses by Students, Parents, and Schools},
volume = {6},
journal = {International Journal of Conflict and Violence}
}

@article{di2024patterns,
  title={Patterns of linguistic simplification on social media platforms over time},
  author={Di Marco, Nicola and Loru, Edoardo and Bonetti, Anita and Serra, Alessandra Olga Grazia and Cinelli, Matteo and Quattrociocchi, Walter},
  journal={Proceedings of the National Academy of Sciences},
  volume={121},
  number={50},
  pages={e2412105121},
  year={2024},
  publisher={National Academy of Sciences}
}

@article{bar2024generative,
  title={Generative AI may backfire for counterspeech},
  author={B{\"a}r, Dominik and Maarouf, Abdurahman and Feuerriegel, Stefan},
  journal={arXiv preprint arXiv:2411.14986},
  year={2024}
}

@inproceedings{schieb2016governing,
  title={Governing hate speech by means of counterspeech on Facebook},
  author={Schieb, Carla and Preuss, Mike},
  booktitle={66th ica annual conference, at fukuoka, japan},
  pages={1--23},
  year={2016}
}

@article{chung2023understanding,
  title={Understanding counterspeech for online harm mitigation},
  author={Chung, Yi-Ling and Abercrombie, Gavin and Enock, Florence and Bright, Jonathan and Rieser, Verena},
  journal={arXiv preprint arXiv:2307.04761},
  year={2023}
}

@misc{instagram-report,
  author       = {{Instagram Help Center}},
  title        = {How to Report Things},
  year         = {2025},
  url          = {https://help.instagram.com/547601325292351},
  note         = {Accessed: 4 September 2025}
}

@article{garland2022impact,
  title={Impact and dynamics of hate and counter speech online},
  author={Garland, Joshua and Ghazi-Zahedi, Keyan and Young, Jean-Gabriel and H{\'e}bert-Dufresne, Laurent and Galesic, Mirta},
  journal={EPJ data science},
  volume={11},
  number={1},
  pages={3},
  year={2022},
  publisher={Springer Berlin Heidelberg}
}

@article{ali2023getting,
  title={Getting meta: A multimodal approach for detecting unsafe conversations within instagram direct messages of youth},
  author={Ali, Shiza and Razi, Afsaneh and Kim, Seunghyun and Alsoubai, Ashwaq and Ling, Chen and De Choudhury, Munmun and Wisniewski, Pamela J and Stringhini, Gianluca},
  journal={Proceedings of the ACM on human-computer interaction},
  volume={7},
  number={CSCW1},
  pages={1--30},
  year={2023},
  publisher={ACM New York, NY, USA}
}

@article{varela2022ignore,
  title={To ignore or not to ignore: The differential effect of coping mechanisms on depressive symptoms when facing adolescent cyberbullying},
  author={Varela, Jorge J and Hern{\'a}ndez, Crist{\'o}bal and Berger, Christian and Souza, Sidclay B and Pacheco, Emanuel},
  journal={Computers in Human Behavior},
  volume={132},
  pages={107268},
  year={2022},
  publisher={Elsevier}
}

@article{benesch2014countering,
  title={Countering dangerous speech: New ideas for genocide prevention},
  author={Benesch, Susan},
  journal={Available at SSRN 3686876},
  year={2014}
}

@article{akhter2022abusive,
  title={Abusive language detection from social media comments using conventional machine learning and deep learning approaches},
  author={Akhter, Muhammad Pervez and Jiangbin, Zheng and Naqvi, Irfan Raza and AbdelMajeed, Mohammed and Zia, Tehseen},
  journal={Multimedia Systems},
  volume={28},
  number={6},
  pages={1925--1940},
  year={2022},
  publisher={Springer}
}

@misc{help_center_2017, author = {Instagram}, title={How to combat bullying and harassment on Instagram},   url={https://help.instagram.com/464473649316860/?helpref=related_articles}, journal={Instagram.com}, year={2017} }

@misc{how_to_handle_x_abuse, author = {X}, title={About online abuse},  url={https://help.x.com/en/safety-and-security/cyber-bullying-and-online-abuse}, journal={help.x.com}, year = {2025}}

@misc{facebook_help_center,author = {Facebook},title={How to handle bullying, harassment, or personal attack on Facebook}, url={https://www.facebook.com/help/116326365118751/}, journal={www.facebook.com}, year = {2025} }

@article{mahbub2021detection,
  title={Detection of harassment type of cyberbullying: A dictionary of approach words and its impact},
  author={Mahbub, Syed and Pardede, Eric and Kayes, ASM},
  journal={Security and Communication Networks},
  volume={2021},
  number={1},
  pages={5594175},
  year={2021},
  publisher={Wiley Online Library}
}

@inproceedings{di2016unsupervised,
  title={Unsupervised cyber bullying detection in social networks},
  author={Di Capua, Michele and Di Nardo, Emanuel and Petrosino, Alfredo},
  booktitle={2016 23rd International conference on pattern recognition (ICPR)},
  pages={432--437},
  year={2016},
  organization={IEEE}
}

@misc{vilk_lo_2023, title={Shouting into the Void: Why Reporting Abuse to Social Media Platforms Is So Hard and How to Fix It}, url={https://pen.org/report/shouting-into-the-void/}, journal={PEN America}, author={Vilk, Viktorya and Lo , Kat}, year={2023}, month={Jun} }

@article{azeez2023classification,
  title={Classification of virtual harassment on social networks using ensemble learning techniques},
  author={Azeez, Nureni Ayofe and Fadhal, Emad},
  journal={Applied Sciences},
  volume={13},
  number={7},
  pages={4570},
  year={2023},
  publisher={MDPI}
}

@inproceedings{anand2019classification,
  title={Classification of abusive comments in social media using deep learning},
  author={Anand, Mukul and Eswari, R},
  booktitle={2019 3rd international conference on computing methodologies and communication (ICCMC)},
  pages={974--977},
  year={2019},
  organization={IEEE}
}

@article{kumbale2023bree,
  title={Bree-hd: A transformer-based model to identify threats on twitter},
  author={Kumbale, Sinchana and Singh, Smriti and Poornalatha, G and Singh, Sanjay},
  journal={IEEE Access},
  volume={11},
  pages={67180--67190},
  year={2023},
  publisher={IEEE}
}

@misc{Detoxify,
  title={Detoxify},
  author={Hanu, Laura and {Unitary team}},
  howpublished={Github. https://github.com/unitaryai/detoxify},
  year={2020}
}

@misc{shahane_2020, title={Cyberbullying Dataset}, url={https://www.kaggle.com/datasets/saurabhshahane/cyberbullying-dataset}, journal={www.kaggle.com}, author={Shahane, Saurabh}, year={2020} }

@article{sanh2019distilbert,
  title={DistilBERT, a distilled version of BERT: smaller, faster, cheaper and lighter},
  author={Sanh, Victor and Debut, Lysandre and Chaumond, Julien and Wolf, Thomas},
  journal={arXiv preprint arXiv:1910.01108},
  year={2019}
}

@article{quintana2022beyond,
  title={Beyond cyberbullying: Investigating when and how cybervictimization predicts suicidal ideation},
  author={Quintana-Orts, Cirenia and Rey, Lourdes and Neto, F{\'e}lix},
  journal={Journal of interpersonal violence},
  volume={37},
  number={1-2},
  pages={935--957},
  year={2022},
  publisher={Sage Publications Sage CA: Los Angeles, CA}
}

@article{nesi2018transformation,
  title={Transformation of adolescent peer relations in the social media context: Part 2—application to peer group processes and future directions for research},
  author={Nesi, Jacqueline and Choukas-Bradley, Sophia and Prinstein, Mitchell J},
  journal={Clinical child and family psychology review},
  volume={21},
  number={3},
  pages={295--319},
  year={2018},
  publisher={Springer}
}

@misc{patchin_2022, title={Summary of Our Cyberbullying Research (2007-2025)}, url={https://cyberbullying.org/summary-of-our-cyberbullying-research}, journal={Cyberbullying Research Center}, author={Patchin, Justin W}, year={2025}, month={Jun} }

@article{nesi2021social,
  title={Social media use and self-injurious thoughts and behaviors: A systematic review and meta-analysis},
  author={Nesi, Jacqueline and Burke, Taylor A and Bettis, Alexandra H and Kudinova, Anastacia Y and Thompson, Elizabeth C and MacPherson, Heather A and Fox, Kara A and Lawrence, Hannah R and Thomas, Sarah A and Wolff, Jennifer C and others},
  journal={Clinical psychology review},
  volume={87},
  pages={102038},
  year={2021},
  publisher={Elsevier}
}

@misc{vogels_2021_1, title={Online Harassment Occurs Most Often on Social media, but Strikes in Other places, Too}, url={https://www.pewresearch.org/short-reads/2021/02/16/online-harassment-occurs-most-often-on-social-media-but-strikes-in-other-places-too/}, journal={Pew Research Center}, author={Vogels, Emily A.}, year={2021}, month={Feb} }

@article{hancock2020ai,
  title={AI-mediated communication: Definition, research agenda, and ethical considerations},
  author={Hancock, Jeffrey T and Naaman, Mor and Levy, Karen},
  journal={Journal of Computer-Mediated Communication},
  volume={25},
  number={1},
  pages={89--100},
  year={2020},
  publisher={Oxford University Press}
}

@inproceedings{hamlett2022labeled,
  title={A labeled dataset for investigating cyberbullying content patterns in instagram},
  author={Hamlett, Mara and Powell, Grace and Silva, Yasin N and Hall, Deborah},
  booktitle={Proceedings of the international AAAI conference on web and social media},
  volume={16},
  pages={1251--1258},
  year={2022}
}

\section{Appendix}
\setcounter{table}{0}
\setcounter{figure}{0}
\renewcommand{\thetable}{A\arabic{table}}
\renewcommand{\thefigure}{A\arabic{figure}}


\begin{table*}[t]
\centering
\setlength{\tabcolsep}{0pt}
\renewcommand{\arraystretch}{1.15}
\begin{tabular}{@{}P{0.96\textwidth}@{}}
\toprule
\textbf{Classification pipeline LLM 1 system prompt} \\
\midrule
You are an expert in discerning authentic online harassment messages on Instagram. You will be given a conversation.

You need to determine if the last message in the given conversation is online harassment that is targeting someone. Someone can be a person in or not in the conversation.

The definition of online harassment is the interpersonal aggression or offensive behavior(s) that is communicated over the internet or through other electronic media.

\vspace{1\baselineskip}

Here are rules you must follow:

1. If the last message looks like online harassment targeting someone, you must reply 1 and provide reasoning.

2. If the last message does not look like online harassment targeting someone, you must reply 0 and provide reasoning.

3. If another message in the conversation looks like online harassment targeting someone, while the last message does not, you must reply 0 and provide reasoning.

4. Give 1 label if the online harassment is targeted at someone.

5. Take the other messages in the conversation into account when labeling the last message. Meanwhile, you only label whether the last message is online harassment targeting someone. 

\vspace{1\baselineskip}

Here are some suggestions to accurately identify online harassment:

1. Passive-aggressive messages are not online harassment.

2. Try not to stretch the meaning of a message.

3. Emojis alone don't carry enough meaning. They can almost never be online harassment.

4. If the message has unintelligible words or phrases, it may have a typo.

\vspace{1\baselineskip}

Here is an additional guideline:

1. In the message, if there are apparently harmful words targeting someone, then it's definitely online harassment.
\\
\bottomrule
\end{tabular}
\caption{Classification pipeline LLM 1 system prompt. Prompt examples constructed
from private conversations are omitted to preserve participant confidentiality.}
\label{tab:a1_system_prompt}
\end{table*}

\begin{table*}[t]
\centering
\setlength{\tabcolsep}{0pt}
\renewcommand{\arraystretch}{1.15}
\begin{tabular}{@{}P{0.96\textwidth}@{}}
\toprule
\textbf{Classification pipeline LLM 1 user prompt} \\
\midrule
You are given an online conversation. You only label whether the last message is online harassment. The last message is marked with ``(label this message)''.

\vspace{1\baselineskip}

The definition of online harassment is the interpersonal aggression or offensive behavior(s) that is communicated over the internet or through other electronic media.

\vspace{1\baselineskip}

Try not to guess if there are unknown terms.

\vspace{1\baselineskip}

Below is an example conversation where the last message is online harassment and you must reply 1.

\vspace{1\baselineskip}

(Three classification examples)

\vspace{1\baselineskip}

If you are sufficiently sure that the last message is online harassment targeting someone, reply ``1''. Provide an one sentence reasoning as well.

If the last message is not online harassment, or if you are not sure, reply ``0''. Provide an one sentence reasoning as well.
You must never use the word imply in your reasoning.

\vspace{1\baselineskip}

Here is the conversation you need to label:

\{csv\_input\}

\vspace{1\baselineskip}

Do you think the last message in the conversation is online harassment targeting someone? 

\vspace{1\baselineskip}

Your label for the last message in the given conversation and reasoning is:
\\
\bottomrule
\end{tabular}
\caption{Classification pipeline LLM 1 user prompt. Prompt examples constructed
from private conversations are omitted to preserve participant confidentiality.}
\label{tab:a1_user_prompt}
\end{table*}

\begin{table*}[t]
\centering
\small 
\setlength{\tabcolsep}{0pt}
\renewcommand{\arraystretch}{1.15}
\begin{tabular}{@{}P{0.96\textwidth}@{}}
\toprule
\textbf{Classification pipeline LLM 2 system prompt} \\
\midrule
You are an expert in discerning authentic online harassment messages on Instagram. You will be given a conversation.

You need to determine if the last message in the given conversation is online harassment that is targeting someone. Someone can be a person in or not in the conversation.

The definition of online harassment is the interpersonal aggression or offensive behavior(s) that is communicated over the internet or through other electronic media.

\vspace{1\baselineskip}

Here are rules you must follow:

1. If you are confident that the last message is online harassment targeting someone, you must reply 1 and provide reasoning.

2. If the last message is not online harassment targeting someone, you must reply 0 and provide reasoning.

3. If another message in the conversation is online harassment targeting someone, while the last message is not, you must reply 0 and provide reasoning.

4. Give 1 label if the online harassment is targeted at someone.

5. Take the other messages in the conversation into account when labeling the last message. Meanwhile, you only label whether the last message is online harassment targeting someone. 

6. You provide label 1 when you are at least kind of sure.

\vspace{1\baselineskip}

Here are some suggestions to accurately identify online harassment:

1. Passive-aggressive messages are not online harassment.

2. Never stretch the meaning of a message.

3. Emojis alone don't carry enough meaning. They can almost never be online harassment.

4. If the message has unintelligible words or phrases, it may have a typo, not online harassment.

\vspace{1\baselineskip}

Here are some additional guidelines:

1. In the message, if there are apparently harmful words targeting someone, then it's definitely online harassment.

2. Do not overthink the tone of the message.

3. Do not overthink how one message implies to be sarcastic.

4. Do not overthink how one message implies to be manipulative.

5. You must never use the word ``imply'' in your reasoning.

6. Generally speaking, online harassment is rare among ordinary conversations.
\\
\bottomrule
\end{tabular}
\caption{Classification pipeline LLM 2 system prompt. Prompt examples constructed
from private conversations are omitted to preserve participant confidentiality.}
\label{tab:a2_system_prompt}
\end{table*}

\begin{table*}[t]
\centering
\small   
\setlength{\tabcolsep}{0pt}
\renewcommand{\arraystretch}{1.15}
\begin{tabular}{@{}P{0.96\textwidth}@{}}
\toprule
\textbf{Classification pipeline LLM 2 user prompt} \\
\midrule
You are given an online conversation. You only label whether the last message is online harassment. The last message is marked with ``(label this message)''.

\vspace{1\baselineskip}

The definition of online harassment is the interpersonal aggression or offensive behavior(s) that is communicated over the internet or through other electronic media.

\vspace{1\baselineskip}

(Three classification examples)

\vspace{1\baselineskip}

If you are absolutely sure that the last message is online harassment targeting someone, reply ``1''. Provide an one sentence reasoning as well.

If the last message is not online harassment, or if you are not absolutely sure, reply ``0''. Provide an one sentence reasoning as well.
You must never use the word imply in your reasoning.

\vspace{1\baselineskip}

Here is the conversation you need to label:

\{csv\_input\}

\vspace{1\baselineskip}

Here is another labeler's label. The first number (0 or 1) is their label, and the following sentence is the reasoning.

\{previous\_result\}

\vspace{1\baselineskip}

The other labeler is just as experienced as you are. Your role is to provide your own independent opinion.

\vspace{1\baselineskip}

Do you think the last message in the conversation is absolutely online harassment targeting someone? 

\vspace{1\baselineskip}

Your label for the last message in the given conversation and reasoning is:
\\
\bottomrule
\end{tabular}
\caption{Classification pipeline LLM 2 user prompt. Prompt examples constructed
from private conversations are omitted to preserve participant confidentiality.}
\label{tab:a2_user_prompt}
\end{table*}




\begin{table*}[!t]
\centering
\footnotesize
\setlength{\tabcolsep}{4pt}
\renewcommand{\arraystretch}{1.1}

\begin{minipage}{0.48\textwidth}
\centering
(a) toxic-bert

\vspace{2pt}

\begin{tabular}{lcccc}
\toprule
\textbf{Class} & \textbf{Prec.} & \textbf{Rec.} & \textbf{F1} & \textbf{Sup.} \\
\midrule
0 & 0.9963 & 0.9820 & 0.9891 & 7485 \\
1 & 0.0878 & 0.3250 & 0.1383 & 40 \\
\midrule
\textbf{Acc.} & \multicolumn{3}{c}{0.9785} & 7525 \\
\textbf{Macro} & 0.5421 & 0.6535 & 0.5637 & 7525 \\
\textbf{Weighted} & 0.9915 & 0.9785 & 0.9846 & 7525 \\
\bottomrule
\end{tabular}
\end{minipage}
\hfill
\begin{minipage}{0.48\textwidth}
\centering
(b) Ensemble

\vspace{2pt}

\begin{tabular}{lcccc}
\toprule
\textbf{Class} & \textbf{Prec.} & \textbf{Rec.} & \textbf{F1} & \textbf{Sup.} \\
\midrule
0 & 0.9968 & 0.9890 & 0.9929 & 7485 \\
1 & 0.1633 & 0.4000 & 0.2319 & 40 \\
\midrule
\textbf{Acc.} & \multicolumn{3}{c}{0.9859} & 7525 \\
\textbf{Macro} & 0.5800 & 0.6945 & 0.6124 & 7525 \\
\textbf{Weighted} & 0.9923 & 0.9859 & 0.9888 & 7525 \\
\bottomrule
\end{tabular}
\end{minipage}

\caption{Classification reports for baseline classifiers on the human-labeled online harassment dataset.}
\label{tab:baseline_classification_reports}
\end{table*}



\begin{table*}[!t]
\centering
\small 
\setlength{\tabcolsep}{0pt}
\renewcommand{\arraystretch}{1.15}
\begin{tabular}{@{}P{0.96\textwidth}@{}}
\toprule
\textbf{AI-based response generation LLM 1 system prompt} \\
\midrule
You are an assistant in helping a user handle online harassment on Instagram. You are given a conversation scenario where the user is sent online harassment by a harasser.

You need to make a decision on how to engage with the harasser based on a list of engagement strategies.

You are helping a teenager around the age of 12\textasciitilde18.

Your goals are (1) helping the user stop the online harassment; (2) deescalating the situation.

In your output, write your choices of engagement strategies with only numbers. Then explain your choices. Your explanation should only be one sentence.
\\
\bottomrule
\end{tabular}
\caption{AI-based response generation LLM 1 system prompt} \label{tab:Simulated_response_Agent_1_system_prompt}
\end{table*}

\begin{table*}[!t]
\centering
\small 
\setlength{\tabcolsep}{0pt}
\renewcommand{\arraystretch}{1.15}
\begin{tabular}{@{}P{0.96\textwidth}@{}}
\toprule
\textbf{AI-based response generation LLM 1 user prompt} \\
\midrule
This conversation happens on Instagram between the user and the harasser:

\{csv\_input\}

\vspace{1\baselineskip}

In the above conversation, the user's messages start with ``User:''. The user is about to respond to the other person in the conversation.

\vspace{1\baselineskip}

You must choose one or multiple of the following engagement strategies to use. You need to choose at least one strategy.

1. Warn the harasser of possible consequences of their actions. \cite{mathew2019thou}

2. Denounce the harasser’s message as being hateful. \cite{mathew2019thou}

3. Establish, maintain, or restore a positive affective relationship with the harasser. \cite{mathew2019thou}

4. Point out the hypocrisy or contradiction in the harasser's messages. \cite{mathew2019thou}

5. Use Empathy to humanize the user and remind the sender that people can be hurt by their behavior. \cite{hangartner2021empathy}

6. Apply moral suasion on the harasser. You may convince the harasser that you are sympathetic and understanding. \cite{munger2021don}

7. Benevolently correct the message with a polite tone, demonstrating understanding and empathy for the harasser. \cite{young2024responding} \cite{reusser2021assessing}

\vspace{1\baselineskip}

In your output, write your choices of engagement strategies with only numbers. Then explain your choices. Your explanation should only be one sentence.
\\
\bottomrule
\end{tabular}
\caption{AI-based response generation LLM 1 user prompt} \label{tab:Simulated_response_Agent_1_user_prompt}
\end{table*}

\begin{table*}[t]
\centering
\small 
\setlength{\tabcolsep}{0pt}
\renewcommand{\arraystretch}{1.15}
\begin{tabular}{@{}P{0.96\textwidth}@{}}
\toprule
\textbf{AI-based response generation LLM 2 system prompt} \\
\midrule
You are an assistant in helping a user respond to online harassment on Instagram.

You will be given a conversation on Instagram where the user is sent online harassment by a harasser.

You will draft 1\textasciitilde2 (more if needed) response messages for the user. Your goals are (1) helping the user stop the online harassment; (2) deescalating the situation. The user will send the messages you wrote consecutively to the harasser.

You are helping a teenager around the age of 12\textasciitilde18.

\vspace{1\baselineskip}

Below are some general engagement strategies you can use when drafting the response messages.

1. Warn the harasser of possible consequences of their actions. \cite{mathew2019thou}

2. Denounce the harasser’s message as being hateful. \cite{mathew2019thou}

3. Establish, maintain, or restore a positive affective relationship with the harasser. \cite{mathew2019thou}

4. Point out the hypocrisy or contradiction in the harasser's messages. \cite{mathew2019thou}

5. Use Empathy to humanize the user and remind the sender that people can be hurt by their behavior. \cite{hangartner2021empathy}

6. Apply moral suasion on the harasser. You may convince the harasser that you are sympathetic and understanding. \cite{munger2021don}

7. Benevolently correct the message with a polite tone, demonstrating understanding and empathy for the harasser. \cite{young2024responding} \cite{reusser2021assessing}

\vspace{1\baselineskip}

You will be given a decision about which of the above strategies to use. You must follow that decision.

\vspace{1\baselineskip}

Below are some writing guidelines you can use when drafting the response messages.

1. Your responses are 1\textasciitilde2 (more if needed) messages. These messages will be sent consecutively to the harasser. Therefore, the messages should look like they are consecutive.

2. Write more than 2 response messages if needed.

3. Use humor when appropriate.

4. Your response messages should be in the tone of a teenager around the age of 12\textasciitilde18. Use the tone and style of the user in the given conversation.

5. Do not sound retaliatory or escalate the situation.

6. Each of your response messages should be around 3 to 13 words.

\vspace{1\baselineskip}

Here are some writing style instructions you should use to sound like a teenager around the age of 12\textasciitilde18.
1. When appropriate, use textese and abbreviations, such as: lol, asap, ikr.
2. When appropriate, you don't need to follow the grammar.
3. When appropriate, use expressive lengthening, such as: sooooo, nooooo, loooool.
\\
\bottomrule
\end{tabular}
\caption{AI-based response generation generation LLM 2 system prompt}
\label{tab:Simulated_response_Agent_2_system_prompt}
\end{table*}

\begin{table*}[t]
\centering
\small 
\setlength{\tabcolsep}{0pt}
\renewcommand{\arraystretch}{1.15}
\begin{tabular}{@{}P{0.96\textwidth}@{}}
\toprule
\textbf{AI-based response generation LLM 2 user prompt} \\
\midrule
This conversation happens on Instagram between the user and the harasser:

\{csv\_input\}

\vspace{1\baselineskip}

In the above conversation, the user's messages start with ``User:''. The user is about to respond to the message sent by the other person in the conversation.

\vspace{1\baselineskip}

Regarding which strategies to use, the decision is: \{previous\_result\}

\vspace{1\baselineskip}

You need to:

1. Draft the 1\textasciitilde2 (more if needed) consecutive response messages that you think would reach your goals.

2. List the strategies used.

3. Present the reasoning behind the response messages.

\vspace{1\baselineskip}

In your output, first write the response messages in separate lines, starting with ``User:''. Then list all the strategies, starting with ``Strategies:'', separating with a comma. Finally, present the reasoning behind the response messages, starting with “Reasoning:”.

Each response message, the strategies part, and the reasoning part should be in separate lines.

Be realistic in the response messages. Do not sound like an AI agent.

Your response messages should be around 3 to 13 words.

For example, an output should be: 

User: Hey.

User: Hello.

Strategies: 1,2,3.

Reasoning: I am responding in this way because they make sense for the scenario.
\\
\bottomrule
\end{tabular}
\caption{AI-based response generation LLM 2 user prompt} \label{tab:Simulated_response_Agent_2_user_prompt}
\end{table*}





\end{document}